**Perspectives on Astrophysics Based on Atomic, Molecular, and Optical (AMO) Techniques**

*Primary Author Information:*
Daniel Wolf Savin
1-212-854-4124
Columbia University
dws26@columbia.edu

*Co-Authors:*
James F. Babb – Harvard-Smithsonian Center for Astrophysics
Paul M. Bellan – California Institute of Technology
Crystal Brogan – National Radio Astronomy Organization
Jan Cami – University of Western Ontario, Canada
Paola Caselli – Max Planck Institute for Extraterrestrial Physics, Germany
Lia Corrales – University of Michigan
Gerardo Dominguez – California State University San Marcos
Steven R. Federman – University of Toledo
Chris J. Fontes – Los Alamos National Laboratory
Richard Freedman – NASA Ames Research Center
Brad Gibson – University of Hull, United Kingdom
Leon Golub – Harvard-Smithsonian Center for Astrophysics
Thomas W. Gorczyca – Western Michigan University
Michael Hahn – Columbia University
Sarah M. Hörst – Johns Hopkins University
Reggie L. Hudson – NASA Goddard Space Flight Center
Jeffrey Kuhn – University of Hawai'i
James E. Lawler – University of Wisconsin Madison
Maurice A. Leutenegger – NASA Goddard Space Flight Center
Joan P. Marler – Clemson University
Michael C. McCarthy – Harvard-Smithsonian Center for Astrophysics
Brett A. McGuire – National Radio Astronomy Organization
Stefanie N. Milam – NASA Goddard Space Flight Center
Nicholas A. Murphy – Harvard-Smithsonian Center for Astrophysics
Gillian Nave – National Institute of Standards and Technology
Aimee A. Norton – Stanford University
Anthony Papol – Brown University
John C. Raymond – Harvard-Smithsonian Center for Astrophysics
Farid Salama – NASA Ames Research Center
Ella M. Sciamma-O'Brien – NASA Ames Research Center
Randall Smith – Harvard-Smithsonian Center for Astrophysics
Chad Sosolik – Clemson University
Clara Sousa-Silva – Massachusetts Institute of Technology
Phillip C. Stancil – University of Georgia Athens
Frank Timmes – Arizona State University
Virginia L. Trimble – University of California Irvine
Bradford J. Wargelin – Harvard-Smithsonian Center for Astrophysics

**A Brief History of Atomic, Molecular, and Optical (AMO) Science and the Next Frontier**

About two generations ago, a large part of AMO science was dominated by experimental high energy collision studies and perturbative theoretical methods. Since then, AMO science has undergone a transition and is now dominated by quantum, ultracold, and ultrafast studies. But in the process, the field has passed over the complexity that lies between these two extremes. Most of the Universe resides in this intermediate region. We put forward that the next frontier for AMO science is to explore the AMO complexity that describes most of the Cosmos.

**Scientific Opportunities in Astrophysics Requiring Advances in AMO Science**

The field of astrophysics advances, in part, through continual improvements in telescopic capabilities, such as collecting area and spatial resolution, and corresponding advances in the associated spectroscopic instrumentation. However, our understanding of the relevant AMO processes that control the observed properties of the Cosmos has not kept pace with these telescopic and spectroscopic improvements. Observations by these new facilities invariably turn up shortcomings in our understanding of AMO physics. As a result, our astronomical explorations are actually driving new theoretical and experimental studies in AMO science.

Many of the new facilities are $1B class investments by the NSF or NASA. Maximizing the scientific return of these facilities hinges, to a large extent, on significant advances in AMO science beyond our current capabilities. Below, we highlight a few of the many astrophysical advances that will become possible through yet-to-be-realized progress in AMO science.

*The molecular Universe:* We have entered a new era of discovery, due to facilities such as the Atacama Large Millimeter/submillimeter Array (ALMA), the Stratospheric Observatory for Infrared Astronomy (SOFIA), and the upcoming *James Webb Space Telescope* (*JWST*). Over 200 molecules have been identified to date in the interstellar medium and in circumstellar media [1]. However, the species responsible for the vast majority of the astronomically observed molecular spectral features remain unidentified [2,3]. Many of these are thought to be due to complex molecules such as polyaromatic hydrocarbons (PAHs), which are abundant in space [4]. Theoretical methods are not sufficiently accurate to correctly predict molecular spectra. Laboratory spectra are the only reliable way at present to identify molecules in space, but the experiments are difficult and time consuming. As a result, new molecules are being identified in space at the rate of about 3.7 per year and it will take decades to catch up with existing observations. *New AMO theoretical and experimental advances in molecular spectroscopy are needed in order to unlock the full mystery of the molecular Universe.*

Molecular observations also reveal important gas-phase chemical processes driving the evolution of baryonic gas, such as inorganic chemistry forming dust (raw material for planets) and organic chemistry leading to the formation of aromatic molecules and other molecules necessary for life as we know it. The complexity of the multi-atom reaction complexes challenges current quantum mechanical and experimental capabilities, most of the relevant gas-phase reaction rate coefficients can only be roughly estimated [5]. *New AMO theoretical and experimental advances in reactive scattering techniques are needed so that we can reliably model and interpret the molecular properties of the Cosmos.*

*Star and planet formation:* Understanding the evolutionary pathway of baryonic matter from atoms in space to the formation of a planetary system around a host star lies at the forefront of current day astrophysical research. This pathway encompasses diffuse atomic and molecular clouds, dense molecular clouds, prestellar cores, protostars, protoplanetary disks, interplanetary dust particles, comets, and meteorites. *New AMO theoretical and experimental advances in various reaction processes are needed to trace out this evolutionary history.*

The transition from diffuse atomic to diffuse molecular clouds can be studied using observations of molecules such as $OH^+$, $CF^+$, $HCl^+$, and $ArH^+$, while observations of the exoergically formed $CH^+$ and $SH^+$ are used to study the input of mechanical energy into the gas [6]. Properly interpreting the astrophysical properties of the observed clouds requires quantitatively understanding the scattering processes that control the observed molecular abundances, in particular dissociative recombination (DR), which destroys these molecules. Accurate DR calculations are challenging, as it is a many-body problem with multiple electrons and nuclei, and common theoretical approximations are usually inadequate, making a rigorous theoretical description of the problem quite challenging, thus requiring new theoretical advances. In the lab, the challenge is to measure DR on heavy cations with internal excitation levels of only about 10 to 100 K. This is likely to become possible with the advent of the heavy ion Cryogenic Storage Ring (CSR) at the Max Planck Institute for Nuclear Physics in Heidelberg, Germany [7].

Studies of star forming clouds use observations of ortho (o-) and para (p-) $H_2D^+$ and $D_2H^+$ to determine the age of the prestellar cores, follow their evolution, and infer the role that magnetic fields may play in the collapse of the cloud [8]. Using the observed ratios of o-$H_2D^+$/p-$H_2D^+$ and o-$D_2H^+$/p-$D_2H^+$, combined with chemical models for the nuclear spin evolution of the gas, estimates can be made for the age of the cloud and compared to the gravitational free-fall time. The inferred ages can exceed that of the free-fall timescale, implying that magnetic fields may be slowing the collapse of the cloud [9]. However, the reliability of these studies is limited by existing uncertainties in the reaction o/p-$H_2$ + o/p-$H_3^+$ → o/p-$H_2$ + o/p-$H_3^+$, and the deuterated isotopic variations, which challenge current quantum theoretical capabilities.

Protoplanetary disks are studied using deuterated molecules. These are powerful probes of cold astrophysical environments, yielding information about the temperature, density, chemistry, abundances, ionization level, evolutionary stage, and thermal history of the disk [10,11]. The diagnostic power of D-bearing molecules arises because the zero-point energy of deuterated molecules is lower than the normal isotope, typically by several hundred kelvin. Important species driving deuteration are $H_2D^+$, $N_2D^+$, $CH_2D^+$, $C_2HD^+$, and $CH_4D^+$ [11,12]. Again, the complexity of these reactive scattering systems requires advances in current quantum mechanical theory and challenges current experimental methods.

*Exoplanetary atmospheres:* Clues to the habitability of exoplanets are provided by their atmospheres [13,14]. As a planet passes before its host star, the star light is filtered by the planet's atmosphere, yielding spectroscopic data. The planned 2021 launch of *JWST* will open up the near- and mid-infrared (IR) range to spectroscopy of planetary atmospheres. But our ability to interpret exoplanetary atmospheres is limited by shortcomings in our understanding of the underlying molecular physics that generates the observed spectra [15]. The spectra of many molecules are incomplete, incorrect, or completely unknown. Some of the important molecules include: $H_2O$, $CO_2$, $CH_4$, $O_3$, CO, $NH_3$, TiO, VO, HCN, $C_2S_2$, $H_2S$, $PH_3$, $SO_2$, HCl, HF, OH, SiO, KOH, and KCl. Data are needed for pressure-induced line broadening parameters, continuum opacity due to collision-induced absorption, molecular opacities at high spectral resolution, photoabsorption cross sections for molecules at high temperatures, and expanded databases for atmospherically relevant chemical reactions. *Meeting these needs will require new intellectual advances in theoretical and experimental molecular physics methods.*

*X-ray astrophysics:* X-ray observations probe some of the most energetic events in the Cosmos. Observations of supernova remnants provide information on the physics of the explosion and the corresponding nucleosynthesis. Active galactic nuclei host supermassive black holes with masses of $10^6$ to $10^9$ $M_{Sun}$. Accretion onto the central black hole feeds energy into the galaxy and regulates star formation. Clusters of galaxies are the most massive gravitationally bound objects in the Universe, and studies of clusters can be used to constrain various cosmological parameters. These are all extended objects and require studies at high spatial and spectral resolution, such as was provided by the X-ray microcalorimeter on the recently lost *Hitomi* mission [16] and will be by the *X-ray Imaging and Spectroscopy Mission* planned for launch in 2021. Interpreting the collected spectra requires an accurate understanding of the underlying atomic physics for elements with atomic number $Z \leq 30$. Important data and processes include: transition wavelengths, rates, and branching ratios; electron impact excitation; innershell excitation and the corresponding Auger and fluorescence yields; electron impact ionization; the electron-ion recombination process known as dielectronic recombination and the corresponding emission line wavelengths and fluxes; and charge balance calculations. *Although these processes have been studied for decades, shortcomings in our AMO understanding still limit our ability to reliably interpret astrophysical observations, often related to the underlying highly-correlated, mult-electron interactions inherent in heavy, highly-ionized species.*

*Nucleosynthesis by kilonovae:* Merging neutron stars were found using gravitational waves and seem to match predictions that they are the formation site for most of the elements with atomic numbers $Z \geq 44$. Optical and infrared spectra from the expanding remnants enable us to study the rapid neutron-capture nucleosynthesis processes that occur during the merger [17]. To this end, opacities are needed for low ionization stages of heavy elements, especially for the lanthanides and actinides. High-level atomic structure methods can provide reasonable atomic data for low resolutions spectral analysis, but it is only in combination with laboratory work and critical evaluations of the data that the needed high resolution spectral information can be generated. *Significant advances are required in both AMO theoretical and experimental methods in order to maximize our scientific understanding of these newly discovered objects.*

*Solar physics:* The recent launch of the *Parker Solar Probe*, the 2019 commissioning of the Daniel K. Inoyue Solar Telescope (DKIST), and the upcoming launch of the *Interstellar Mapping and Acceleration Probe* all herald a new era in solar physics. These facilities and missions will study the magnetohydrodynamics and dynamo processes of the Sun, flares and eruptive events that can affect life on Earth, the flow of mass and energy through the solar atmosphere, and the long-term behavior of the Sun. Of particular relevance to AMO science are the DKIST spectroscopic capabilities and corresponding needs [18]. The DKIST spectral range will extend from the visible into the near-IR, the latter of which is largely unexplored spectroscopically. Line identifications, oscillator strengths, transition rates, and branching ratios will be needed for multi-electron systems. New magnetic field diagnostics in the IR will also become available using the Hanle and Zeeman effects, all of which will require the relevant atomic data. Lastly, the solar plasma is dynamic and not in equilibrium. The ionization states of abundant elements encode the thermal history of the plasma, but they can only be interpreted in terms of plasma heating processes if accurate ionization and recombination data are known. *Advances in AMO theory and experiment are needed to generate all of the above required data.*

**AMO Contributions to new Tools for Astrophysics**

*Exoplanets:* Over 4,000 exoplanets have been discovered to date. Observational constraints have limited these primarily to planets more massive than Earth or with smaller separations from their host stars. Most planets are unlikely to transit across their host stars from our perspective, precluding the use of eclipse methods. Hence, future searches for Earth-like planets will be dominated by detecting Doppler shifts in the stellar spectrum induced by the orbital motion of the planet. These studies will require visible and IR spectrometers calibrated to an accuracy of a few parts in $10^9$ or better and with a stability on the order of decades. Commercially available Th/Ar calibration lamps are contaminated by ThO, iodine lamps affect the measured stellar spectra, and thus new alternatives are needed, with U/Ne being one such proposed lamp. The use of laser frequency combs or Fabry-Perot sources has been proposed, but the long-term stability of these are untested. *Advances in AMO science would revolutionize our ability to detect exoplanets with masses and orbits similar to Earth's.*

*Gravitational waves:* Squeezed light methods, combined with enhanced power in the interferometer detectors, are predicted to enhance gravitational wave detections at high frequencies (50 – 5,000 Hz). These advances will double the spatial detection volume, provide better sky localization, enable better estimates for the tidal deformability parameters of the merging compact objects, open up the ability to study the post-merger phase, and constrain the neutron star equation of state. *Continued advances in AMO techniques hold the promise of expanding our ability to use gravitational waves to help unravel the mystery of the Cosmos.*

### Future AMO Science Enabled by Astrophysics

*Fundamental constants:* Astronomical spectra can be used to test for variations in fundamental constants such as the fine structure constant α and the ratio of the electron-to-proton mass μ [19]. *These studies require highly accurate atomic and molecular structure data that will require new intellectual advances in AMO scientific capabilities.*

### Workforce, Funding, and Facilities

*Workforce:* There are very few AMO faculty working to address the needs of the astrophysics community. This area is referred to as AMO laboratory astrophysics, and encompasses all theoretical and experimental studies relevant to astrophysics. Physics Departments are typically interested in quantum, ultracold, and ultrafast studies and point AMO laboratory astrophysicists to Astronomy Departments. On the other hand, Astronomy Departments are primarily interested in observational and astrophysical theoretical/modeling studies. So if one is studying molecules, Astronomy Departments will direct that person to Chemistry Departments. But Chemistry Departments are primarily focused on large molecules and biochemistry and are uninterested in astrophysically important molecules such as PAHs or C- and Si-based structures. Smaller molecules, they say, are the realm of a Physics Department. But traditional AMO laboratory are designed primarily for the study of two- or three-atom molecules and are poorly equipped to study more complex systems. So AMO laboratory astrophysics lies at the interface of three academic departments, none of whom want to take ownership. In addition, the workforce issue begins early in the academic pipeline with junior scientists finding it difficult to get professional and financial support for their laboratory astrophysics work. These issues raise the serious question: how will we train the next generation of AMO laboratory astrophysicists? Mirroring this issue, the National Academy of Sciences also lacks the robust leadership in this critical interdisciplinary area that is needed to advance the field.

*Funding:* It is extremely hard to get funding for AMO laboratory astrophysics. This appears to be due to a narrow vision by the NSF, Department of Energy (DOE), and their review panels. One proposal submitted to the NSF Division of Physics Experimental AMO program was declined, in part, because "This work is more appropriately funded through the Foundation's Division of Astronomy [sic] or by NASA." What the agency and reviewers continually fail to recognize is that the solution for many of the issues lies squarely in the AMO intellectual span. Once the solution has been developed, the techniques can then be exported to the astrophysics community, and support can be obtained from astrophysics programs. Another NSF proposal submitted to the same program was declined, in part, because "There does not seem to be a pre-existing demand from theory for the measurements." It is bad for AMO science if everyone is only doing research that theory is already studying. Expanding the frontiers of knowledge means going where others have not been before. Lastly, DOE support for AMO laboratory astrophysics research has dried up over the last few decades and does not appear likely to be resurrected.

*Facilities:* AMO laboratory astrophysics facilities range in scale from table top to national laboratory. Table-top devices, excellent for training students, are disappearing as AMO laboratory astrophysics faculty retire and close their labs. The faculty are not being replaced. For the soft-money scientists who remain in the field, it is extremely hard to build a new lab without start-up funds. National laboratories can sometimes carry out critically needed studies. For example, the Linac Coherent Light Source – II (LCLS II) will have the wavelength stability to study the opacity of multiply ionized heavy elements needed for kilonova spectra. However, it is extremely hard to get beamtime for a project that is unlikely to result in a Science, Nature, or Phys. Rev. Lett. publication. Supercomputer support and funding for modern code development are also critical for the advanced theoretical computations that are now needed.

## Recommendations

AMO science is the Rosetta stone that helps us to unlock the mysteries of the Cosmos. Exploring the AMO complexity that drives much of the Universe is the next frontier for AMO science. Major astrophysical advances will come from these AMO studies, but the initial intellectual efforts fall solidly within AMO science. Exploration of this AMO complexity requires a rejuvenation of AMO laboratory astrophysics faculty and a corresponding restoration of funding from NSF and DOE for such studies.

## Acknowledgements

The authors thank J. Berengut, P. Bernath, D. Helfand, D. Kipping, J. Kuhn, B. Metzger, T. Oka, Y. Ralchenko, F. Salama, R. Smith, and M. Tse for stimulating discussions.